\documentclass[12pt]{article}

\usepackage{graphicx}
\usepackage{slashed}
\begin{document}

\title{Cogeneration of Dark Matter and Baryons by Non-Standard-Model Sphalerons
in Unified Models}
\author{{\bf S.M. Barr} and {\bf Heng-Yu Chen} \\
Department of Physics and Astronomy \\ 
and \\
Bartol Research Institute \\ University of Delaware
Newark, Delaware 19716} \maketitle

\begin{abstract}
\noindent Sphalerons of a new gauge interaction can convert a primordial asymmetry in $B$ or $L$ into a dark matter asymmetry. From the equilibrium conditions for the sphalerons of both the electroweak and the new interactions, one can compute the ratios of $B$, $L$, and $X$, where
$X$ is the dark matter number, thus determining the mass of the dark matter particle
fairly precisely. Such a scenario can arise naturally in the context of unification with larger groups. An illustrative model embeddable in $SU(6) \times SU(2) \subset E_6$
is described in detail as well as an equally simple model based on $SU(7)$. 
\end{abstract}

\section{Introduction} 

The fact that the cosmic densities of dark matter and ordinary baryonic matter
are comparable \cite{OmegaDM} has suggested to many authors that 
they may have a common origin, that is that the dark matter and baryonic 
matter may have been generated by the same processes, or that one
of them may have been generated from the other. This idea is
sometimes called ``cogeneration" of dark matter and ordinary matter. 
There is a
rapidly growing literature studying various ways that this might
have happened
\cite{bcf,adm-hdops,adm-sphaleron,adm-baryonic,adm-other,adm-darksph}.

The first papers to propose this possibility \cite{bcf} were based
on the idea that primordial asymmetries in baryon and lepton number
($B$, $L$) were partially converted into an asymmetry in some other
global quantum number (call it $X$) by sphaleron processes
\cite{sphaleron} when the temperature of the universe was above the
weak interaction scale. Assuming $X$ to be conserved (or
nearly so) at low temperatures, the lightest particles carrying this
quantum number would be stable and could play the role of dark
matter. What would result from such a scenario is ``asymmetric dark
matter" \cite{nussinov}. Many other scenarios for generating
asymmetric dark matter have been proposed
\cite{adm-hdops,adm-sphaleron,adm-baryonic,adm-other}. In some of
these scenarios ordinary matter and dark matter are converted into
each other by perturbative processes involving higher-dimension
operators \cite{adm-hdops}; and in others by sphalerons (or by both
sphalerons and higher-dimension operators) \cite{adm-sphaleron}. In
some scenarios, the dark matter carries baryon number which
compensates for the non-zero baryon asymmetry of ordinary matter
\cite{adm-baryonic}. And many papers propose still other mechanisms
\cite{adm-other}.

A very interesting idea first proposed in \cite{adm-darksph} is that a primordial 
asymmetry in $B$ or $L$ (or both)  
is partly converted into an $X$ asymmetry (and thus a dark matter asymmetry)
by sphalerons of some new non-abelian gauge interaction. In this paper we point out
that this mechanism arises very naturally in grand unified models.
In a previous paper \cite{Barr2012}, it was noted by one of us that grand unified models with groups larger than $SU(5)$ provide a natural context for the emergence of dark matter. The larger fermion multiplets of such models typically contain fermions that are Standard Model singlets, which could play the role of dark matter. Unified models can also have accidentally conserved global charges (analogous to $B-L$ in $SU(5)$ models) that could be the charge $X$ carried by dark matter. It was also noted in \cite{Barr2012} that larger unification groups can have additional non-abelian subgroups whose sphalerons could convert $B$ and $L$ asymmetries into an $X$ asymmetry. Here we show that simple models can indeed be constructed that realize this possibility. Most of this paper is devoted to an example
based on an $SU(5) \times SU(2)$ that is embeddable in $E_6$. At the end of the paper we note that a similar and equally simple model can be constructed based on $SU(7)$.
These models exploit all the features of grand unification favorable to the genesis of dark matter that were emphasized in \cite{Barr2012}.

In the model that we present here, we denote by $SU(2)_*$ the gauge group of the new interaction whose sphalerons are responsible for cogenerating the dark matter, in order to distinguish it from the electroweak group $SU(2)_L$. 
An interesting feature of the models we discuss
is that the present ratio of the number densities of dark matter particles and baryons,
$n_{DM}/n_B$, can be calculated thermodynamically from just the particle content of the model
and is independent of the nature of the primordial asymmetry. This allows one to determine the mass of the dark matter particle, which is given simply by $m_{DM} = m_p \frac{\Omega_{DM}}{\Omega_B}
\frac{n_B}{n_{DM}}$.

A general point that is worth noting here is that the task of constructing realistic models is considerably simpler
if the spahalerons that produce the $X$ asymmetry are those of a new non-abelian symmetry 
as in \cite{adm-darksph} rather than those of the electroweak interactions as in \cite{bcf}. In the latter case, there must 
exist fermions that are chiral under $SU(2)_L$ and have $X \neq 0$. Those particles 
cannot be lighter than $M_W$ or they would have been discovered already (since some 
of them have to be electrically charged). But if they are heavy compared to $M_W$ 
they tend to contribute excessively to the $\rho$ parameter. By contrast, if the 
sphalerons that produce the $X$ asymmetry are those of a new non-abelian interaction, then all $X \neq 0$ fermions can be neutral with respect to the Standard Model gauge interactions, thus avoiding the above phenomenological problem.

\section{The Model}

The model we propose is based on the gauge group $G_{SM} \times SU(2)_*$, which can 
be embedded in larger groups in the following way:

\begin{equation}
E_6 \; \supset \; SU(6) \times SU(2) \; \supset \; SU(5) \times SU(2)
\; \supset \; G_{SM} \times SU(2).
\end{equation}

\noindent Moreover, the particle content of the model is exactly what would arise 
from such an embedding. In particular, each family of fermions consists of the 
27 particles that make up the fundamental representation of $E_6$:

\begin{equation}
\begin{array}{l}
{\bf 27} \; \longrightarrow \; (15, 1) + ( \overline{6}, 2) \; \longrightarrow \;
(10, 1) + (5,1) + (\overline{5}, 2) + (1,2) 
\\ \\
\longrightarrow \left( \ell^c, \left[ \begin{array}{c} u \\ d \end{array} \right], u^c  \right) + 
\left( \left[ \begin{array}{c} \overline{\ell} \\ \overline{\nu} \end{array}
\right],  \overline{d^c} \right) + 
\left( \left[ \begin{array}{c} \ell_I \\ \nu_I \end{array}
\right],  d^c_I \right) + (\chi_I), \;\;\;\; I=1,2,
\end{array}
\end{equation}

\noindent where the decomposition in Eq. (2) corresponds to the sequence of groups
in Eq. (1). The index $I$ in Eq. (2) stands there and through the paper for the index of the extra $SU(2)_*$ group. Note that each family automatically contains particles, denoted $\chi_I$, that are singlets under $G_{SM}$ but non-singlet under $SU(2)_*$ and thus able to play the role of dark matter, illustrating the point made in \cite{Barr2012}.

While Eqs. (1) and (2) show our model is naturally unified
in a larger group, this is not essential to the mechanism of cogeneration. 
Henceforth in this paper we will discuss the model as if its gauge group is just $G_{SM} \times SU(2)_*$ without any assumption about whether this is unified at some high scale. Nevertheless, it is convenient as a notational ``shorthand" to refer to fermions and scalars by the 
$SU(5) \times SU(2)_*$ multiplets in which they {\it would} be contained if the model 
were further unified, and we shall often do this.

In $SU(5) \times SU(2)_*$ language, then, each family consists of
$(10,1) + (5, 1) + (\overline{5}, 2) + (1,2)$. Besides the $(1,2) = \chi_I$ already mentioned, there are other non-Standard Model fermions contained in each family,
namely the and {\it half} of the fermions in the $(\overline{5},2)$, or, in terms of $SU(5)$ alone, a vectorlike $5 + \overline{5}$ pair. We will henceforth call these the ``extra fermions". 

The spontaneous breaking of $SU(2)_*$, at a scale $M_*$, is accomplished by the vacuum expectation value (VEV) of a $(1,2)$ multiplet of Higgs fields that we shall denote $\Omega_I$. This Higgs field also gives mass to the ``extra fermions" by means of the following Yukawa coupling 

\begin{equation}
(\overline{5}, 2) \; (5,1) 
\; \langle (1,2)_h \rangle \; \supset \; d^c_I \overline{d^c} \; \langle \Omega_I \rangle, \;\;\;\;   \left(  
\ell_I \;\; \nu_I \right) \cdot \left( \begin{array}{c}
\overline{\ell} \\ \overline{\nu} \end{array} \right) \; \langle \Omega_I \rangle.
\end{equation}

It was said above that the fermions $\chi_I$ that transform as $(1,2)$ play the role of
dark matter. But more precisely, there are three families of these $SU(2)_*$ 
doublets, or altogether six flavors of them, and it is the lightest of them 
that is stable and composes the dark matter. To give these six fermions mass we introduce
six partners for them that are singlets under all the gauge groups. We denote these by $\chi^c_a$, $a = 1,..,6$. 
The Yukawa terms that give them mass are of the form $Y_a (\chi_I \chi^c_a) \langle \Omega_I
\rangle$. 

The value of the scale $M_*$ at which $SU(2)_*$ is broken by $\langle \Omega_I \rangle$ does not matter very much as far as the mechanism for generating a dark matter asymmetry is concerned. It should certainly be large enough that the $SU(2)_*$ gauge bosons and the ``extra" fermions in $(5,1)$ and $(\overline{5}, 2)$ would not already have been detected. On the other hand, the dark matter particles, which will later be seen to have mass around 1 GeV,   
obtain mass from the VEV of $\Omega_I$. Therefore, the larger the VEV of $\Omega_I$ is, the smaller must be its Yukawa coupling to the dark matter particles. We know that some Yukawa couplings in nature are very small (those of $e$, $u$, and $d$ are of order $10^{-5}$).
If one does not wish Yukawa couplings to be smaller than $10^{-5}$, say, one would need
$M_*$ to be less than about $100$ TeV. We imagine, therefore, that $M_*$ is somewhere between 1 and $100$ TeV. Moreover, as will be discussed later, if $M_*$ is larger than about
100 TeV, the $SU(2)_*$ gauge
interactions will be too slow to keep the ``dark sector" of particles in thermal equilibrium with the Standard Model particles long enough to avoid problems with primordial nucleosynthesis (The energy trapped in massless particles of the dark sector can cause the universe too expand too rapidly in the era of primordial nucleosynthesis).

It should be noted that since the $\Omega_I$ is in a pseudo-real 
representation of the gauge group and since we will not give it any global
charge that would distinguish from its conjugate $\Omega^*_I$, the symmetries of the model allow 
$\Omega^*_I$ to couple in the same ways that $\Omega_I$ can. For example, there are both  
$(\chi_I \chi^c_a) \langle \Omega_I
\rangle$ and $(\chi_I \chi^c_a) \langle \Omega^*_I
\rangle$ Yukawa terms, and similarly there are both 
$h h'_I \Omega_I$ and $h h'_I \Omega^*_I$ terms in the Higgs potential. (These facts imply the chemical potential of the $\Omega$ fields zero, which is relevant to our later discussion.) 

To break the electroweak gauge group and give mass to all the Standard Model 
quarks and leptons, there must be 
more than one $SU(2)_L$ doublet of Higgs fields. The masses of the
up quarks come from a Higgs doublet, which we shall denote $h$, that would
be contained in $(5,1)$ of $SU(5) \times SU(2)_*$. In that language, it has  
Yukawa couplings of the type $(10,1)(10,1) \langle (5,1)_h \rangle$, which
contains in particular $\; u u^c \langle h\rangle$.
The down quarks and charged leptons obtain mass from a pair of Higgs 
doublets, which we denote $h'_I$, that would be contained in $(\overline{5}, 2)$
of $SU(5) \times SU(2)_*$. These have Yukawa couplings of the form 
$(10,1)(\overline{5}, 2) \langle (\overline{5}, 2)_h \rangle$, which contains in particular 
$d d^c_I  \langle h'_I \rangle$ and $\ell^c \ell_I \langle h'_I \rangle$. 
The neutrinos can obtain mass from the dimension-5 effective operator
$\nu_I \nu_J \langle h'_I \rangle \langle h'_J \rangle/M_R$. It does not matter for our purposes whether this operator arises from the Type I or Type II see-saw mechanism.

Finally, two more types of particle are contained in the model: some number 
($p$) of gauge singlet fermions that will be denoted $S$ and a gauge singlet
boson that will be denoted $\sigma$. These will play a role, as will be seen, 
in allowing dark matter particles and their antiparticles to annihilate leaving
only ``asymmetric dark matter". 

\newpage

\noindent {\large\bf Table I:} The fermion and scalar content of the model.

\vspace{0.2cm}

\begin{tabular}{|lc|c|c|c|c|c|c|}
\hline
& name & $G_{SM} \times SU(2)_*$ & $SU(5) \times SU(2)_*$ & $SU(6) \times SU(2)_*$ &
$E_6$ & X & W 
\\ \hline
$3 \times$ & $\left( \begin{array}{c} u \\ d \end{array} \right)$ & $(3,2, +\frac{1}{6};1)$ &
$(10,1)$ & $(15,1)$ & $27$ & $0$ & $0$ \\
$3 \times$ & $u^c$ & $(\overline{3},1, -\frac{2}{3};1)$ &
$(10,1)$ & $(15,1)$ & $27$ & $0$ & $0$ \\
$3 \times$ & $\ell^c$ & $(1,1, +1;1)$ &
$(10,1)$ & $(15,1)$ & $27$ & $0$ & $0$ \\
$3 \times$ & $\left( \begin{array}{c} \overline{\ell} \\ \overline{\nu} \end{array} \right)$ & $(1,2, +\frac{1}{2};1)$ &
$(5,1)$ & $(15,1)$ & $27$ & $0$ & $0$ \\
$3 \times$ & $\overline{d^c}$ & $(3,1, -\frac{1}{3};1)$ &
$(5,1)$ & $(15,1)$ & $27$ & $0$ & $0$ \\ \hline
$3 \times$ & $\left( \begin{array}{c} \ell_{1,2} \\ \nu_{1,2} \end{array} \right)$ 
& $(1,2, -\frac{1}{2};2)$ &
$(\overline{5},2)$ & $(\overline{6},2)$ & $27$ & $0$ & $0$ \\
$3 \times$ & $d^c_{1,2}$ & $(\overline{3},1, +\frac{1}{3};2)$ &
$(\overline{5},2)$ & $(\overline{6},2)$ & $27$ & $0$ & $0$ \\
$3 \times$ & $\chi_{1,2}$ & $(1,1, 0;2)$ &
$(1,2)$ & $(\overline{6},2)$ & $27$ & $+1$ & $0$ \\ \hline
 & $\chi^c_{1,...,6}$ & $(1,1,0;1)$ & $(1,1)$ & $(1,1)$ & $1$ & $-1$ & $0$ \\
$p \times$ & $S$ & $(1,1,0;1)$ & $(1,1)$ & $(1,1)$ & $1$ & $0$ & $+1$ \\
\hline 
& $\Omega_{1,2}$ & $(1,1,0;2)$ & $(1,2)$ & $(\overline{6},2)$ & $27$ & $0$ & $0$ \\
 & $h$ & $(1,2,+\frac{1}{2}; 1)$ & $(5,1)$ & $(15,1)$ & $27$ & $0$ & $0$ \\
 & $h'_{1,2}$ & $(1,2,-\frac{1}{2}; 2)$ & $(\overline{5},2)$ & $(\overline{6}, 2)$
& $27$ & $0$ & $0$ \\
& $\sigma$ & $(1,1,0;1)$ & $(1,1)$ & $(1,1)$ & $1$ & $+1$ & $-1$ \\ \hline
\end{tabular}
   
\vspace{1cm}
The complete fermion and scalar content of the model is displayed in Table I. In the last two columns of Table I, we give the charges of the fields under two global symmetries, $U(1)_X$ and $U(1)_W$. The charge $X$ is the crucial one for the model. It is the asymmetry in $X$ that is responsible for the existence of stable dark matter. The charge $W$ plays the role of constraining the couplings of the singlet fields $S$ and $\sigma$ that are responsible for the annihilation of dark matter particles with their anti-particles. In particular, the global $U(1)_W$ invariance means that these fields interact only by the Yukawa term $y(\chi^c S) \sigma$. This term allows the annihilation process $\chi^c + \overline{\chi^c} \longrightarrow S + \overline{S}$ to occur by the exchange of a $\sigma$ boson in the $t$ channel. The $\sigma$ boson is assumed to have no vacuum expectation value, and therefore the $S$ fermions are massless. In this way, essentially all the dark matter anti-particles annihilate with dark matter particles into massless particles, whose energy is red-shifted away as the universe expands, leaving only the dark matter particle excess, i.e. the ``asymmetric dark matter". The global symmetries $U(1)_X$ and $U(1)_W$ can arise as accidental symmetries of the low energy theory even if $G_{SM} \times SU(2)_*$ is unified in a larger group, as will be discussed later.

\section{The Genesis of the Dark Matter Asymmetry}
                            
Now that the particle content and couplings of the model have been defined, we turn
to the process by which the dark matter asymmetry is generated. The sphalerons of $SU(2)_*$ create one each of every left-handed fermion that is a doublet of $SU(2)_*$, namely
(for each family) the three colors of $d^c_I$, the leptons $\nu_I$ and $\ell_I$, and the $X$-bearing particles $\chi_I$.
Thus, for the $SU(2)_*$ sphaleron processes $\Delta X =  \Delta B = \frac{1}{2} \Delta L$.
(We follow the loose but common practice of referring to processes that involve the anomaly of some group $G$ as ``sphalerons" even if they happen at a temperature far above the scale at which $G$ is broken rather than through tunneling.) The sphalerons of the electroweak $SU(2)_L$ give $\Delta X = 0$ and $\Delta B = \Delta L$. 
All other processes at low energy conserve $B$, $L$, and $X$. (There might be grand unified interactions that violate these quantum numbers, and such interactions might have played a role in generating a primordial asymmetry in one or more of them. But when the temperature is far below the grand unification scale, we can neglect these interactions.)

There are four cosmological eras that need to be considered: (a) The era when some primordial asymmetry of $B$, $L$, or $X$ (or some combination of them) was generated.
This could have been by means of grand unified interactions; but in any case we assume that it happened when the temperature was much higher than the scale $M_*$ at which the $SU(2)_*$ interactions are broken. It does not matter for us which of the many mechanisms that have been proposed for baryogenesis or leptogenesis is responsible for this, or what the relative values were of the asymmetries in $B$, $L$, and $X$ that were produced in this primordial era. (b) The era after the primordial asymmetries were generated, but when the temperature is still greater than $T_*$, where $T_*$ is the temperature below which the $SU(2)_*$ sphalerons processes effectively cease. (c) The era when $T_W < T < T_*$, where $T_W$ is the temperature below which the $SU(2)_L$ sphaleron processes effectively cease. ($T_W$ has been estimated to be about 200 GeV \cite{T-Sph}.) And (d), the era when $T< T_W$. 

In era (b), when both kinds of sphaleron processes ($SU(2)_*$ and $SU(2)_L$) are active, the ratios of $X$, $B$, and $L$ are determined by thermodynamics. The point is that the requirement of equilibrium for the two kinds of sphaleron processes gives two conditions on the two independent ratios of these quantum numbers. At the end of era (b), when $T$ falls below $T_*$ and the $SU(2)_*$ sphaleron processes effectively cease, the ratio of $X$ to $B-L$ is frozen, because all other processes conserve both $B-L$ and $X$.

In the next era, when $T_W < T < T_*$, the $SU(2)_L$ sphaleron processes continue to violate $B$ and $L$, and the ratio of $B$ to $L$ changes to a new value that can be computed from the requirement that the $SU(2)_L$ sphalerons are in equilibrium. The ratio of $B$ to $L$ becomes frozen when the temperature falls below $T_W$, since after that point all processes conserve both $B$ and $L$. Consequently, from that point on, down to the present, $X$, $B$, and $L$ remain in constant ratios to each other. 

Finally, when the temperature falls below the mass of the lightest $X$-bearing particle (which is the dark matter particle), virtually all the particles with non-zero $X$ annihilate with their anti-particles into massless $S$ fermions, except the residue that cannot annihilate due to the asymmetry in $X$. (We assume the dark matter annihilation 
is sufficiently fast to leave almost purely asymmetric dark matter. This puts a constraint on the mass and coupling of the scalar $\sigma$, which will be discussed later.) 

We now turn to the thermodynamic calculation of the ratios of $X$, $B$, and $L$, which parallels the calculations in \cite{HarveyTurner}. We start with era (b) when there is already some primordial asymetry and when $T> T_*$. We assume that in this era the $SU(2)_*$ symmetry may be treated as unbroken. Thus all the particles within an irreducible multiplet of 
$G_{SM} \times SU(2)_*$ all have equal chemical potentials, and the chemical potentials of the gauge bosons vanish. Moreover, the scattering processes involving the Yukawa interactions and scalar self-interactions, which we assume to be in equilibrium, give relations among the chemical potentials that allow one to write all of them in terms of just five, namely $\mu_Q$, $\mu_L$, $\mu_{\chi}$, $\mu_h$, and $\mu_{\sigma}$, which are respectively the chemical potentials of the quark doublets $(u, d)$, the lepton doublets $(\nu_I, \ell_I)$, the $\chi_I$, and the scalar fields $h$ and $\sigma$. In particular, we have

\begin{equation}
\begin{array}{l}
\mu_{\Omega} = \mu_{\Omega^*} \; \;  \Rightarrow \;\;   \mu_{\Omega} = 0, \\
\mu_L + \mu_{\overline{L}} + \mu_{\Omega} = 0 \;\; \Rightarrow \;\; 
\mu_{\overline{L}} = - \mu_L, \\
\mu_h + \mu_{h'} + \mu_{\Omega} = 0 \;\;\ \Rightarrow \;\;  \mu_{h'} = - \mu_h, \\
\mu_{u^c} = - \mu_Q - \mu_h, \\
\mu_{d^c} = - \mu_Q + \mu_h, \\
\mu_{\ell^c} = - \mu_L + \mu_h, \\
\mu_{d^c} + \mu_{\overline{d^c}} + \mu_{\Omega} = 0 \;\; \Rightarrow \;\; 
\mu_{\overline{d^c}} = - \mu_{d^c} = \mu_Q - \mu_h, \\
\mu_{\chi} + \mu_{\chi^c} + \mu_{\Omega} = 0 \;\; \Rightarrow \;\; \mu_{\chi^c}=- \mu_{\chi}, \\ 
\mu_{\chi^c} + \mu_S + \mu_{\sigma} = 0 \;\; \Rightarrow \;\; \mu_S = 
\mu_{\chi} - \mu_{\sigma}.
\end{array}
\end{equation}

The next step is to realize that the electric charge $Q$ and the global charge $W$ are conserved by all interactions, and therefore the conditions $Q =0$ and $W=0$ must be satisfied. These two conditions will allow
us to solve for the chemical potentials of the scalars, $\mu_h$ and $\mu_{\sigma}$, in terms of those of the fermions,
$\mu_Q$, $\mu_L$, and $\mu_{\chi}$. In computing the density of $Q$ and $W$, we assume that all the 
particles of the model have masses small compared to $M_*$ and thus to $T$. (When $T \sim T_*$, the mass of $\Omega$ may perhaps not be negligible compared to $T$, but this will not matter for what follows since $\mu_{\Omega} = 0$.) 

The condition for electric charge to vanish is then

\begin{equation}
\begin{array}{ccccl}
0 & = & Q & \propto  & 6 \mu_L (-1) + 3 \mu_{\ell^c} (+1) + 3 \mu_{\overline{L}} (+1) \\ & & & & \\
& & & + & 9 \mu_Q (+ \frac{2}{3}) + 9 \mu_{u^c} (-\frac{2}{3}) + 9 \mu_Q (- \frac{1}{3})
+ 18 \mu_{d^c} (+ \frac{1}{3}) + 9 \mu_{\overline{d^c}} (- \frac{1}{3}) \\ & & & & \\
& & & + & (b(0)/f(0)) [\mu_h (+1) + 2 \mu_{h'} (-1)] \\ & & & & \\
& \Rightarrow & 0  &  = & - 12 \mu_L + 24 \mu_h \;\; \Rightarrow \;\; \mu_h = \frac{1}{2} \mu_L
\end{array}
\end{equation}

\noindent where $f(x) \equiv \frac{1}{4 \pi^2} \int_0^{\infty} y^2 dy 
[\cosh^2 (\frac{1}{2} \sqrt{y^2 + x^2})]^{-1}$ and $b(x) \equiv \frac{1}{4 \pi^2} \int_0^{\infty} y^2 dy [\sinh^2 (\frac{1}{2} \sqrt{y^2 + x^2})]^{-1}$ are integrals over the Fermi and Bose distribution functions and $x = m/T$. Since we are assuming that the 
particle masses are small compared to $T_*$, we have that $b(x)/f(x) \cong b(0)/f(0) = 2$. In obtaining the last line of Eq. (5), we have used the relations given in Eq. (4). In a similar way we have, from the vanishing of $W$,

\begin{equation}
\begin{array}{ccl}
0 & = & p \mu_S (+1) + (b(0)/f(0)) \mu_{\sigma} (-1)     
\;\;\;\; \Rightarrow \;\; \mu_S = \frac{2}{p} \mu_{\sigma} \\ & & \\
& \Rightarrow &  \mu_{\chi} - \mu_{\sigma} = \frac{2}{p} \mu_{\sigma} \;\;\; 
\Rightarrow \;\;  \mu_{\sigma} = \left( \frac{p}{p+2} \right) \mu_{\chi},
\end{array}
\end{equation}

\noindent where to get the last line of Eq. (6), we have used the last relation in Eq. (4).
We remind the reader that the integer $p$ in Eq. (6) is the the number of massless $S$ fields. (See Table I.) The minimal model would therefore simply have $p=1$. 

The final step in analyzing era (b), is to use the equilibrium conditions for the two types of sphalerons to relate the chemical potentials of the fermions, $\mu_Q$, $\mu_L$, and $\mu_{\chi}$. The condition for equilibrium of the $SU(2)_L$ sphalerons is simply

\begin{equation}
0 = 9 \mu_Q + 6 \mu_L + 3 \mu_{\overline{L}} \;\; \Rightarrow \;\; 
\mu_Q = - \frac{1}{3}  \mu_L,
\end{equation}

\noindent where we have used $\mu_{\overline{L}} = - \mu_L$ from Eq. (4). For the $SU(2)_*$ sphalerons, the equilibrium condition is 

\begin{equation}
\begin{array}{ccl}
0 & = & 6 \mu_L + 9 \mu_{d^c} + 3 \mu_{\chi} = 6 \mu_L + 9 (- \mu_Q + \mu_h) 
+ 3 \mu_{\chi}  \\ & & \\
\Rightarrow & 0 & = \frac{21}{2} \mu_L - 9 \mu_Q + 3 \mu_{\chi} \;\; 
\Rightarrow \;\; 0 = \frac{27}{2} \mu_L + 3 \mu_{\chi} \\ & & \\
\Rightarrow & \mu_{\chi} & =  - \frac{9}{2} \mu_L, 
\end{array}
\end{equation}

\noindent where in the middle steps in Eq. (8) we have used Eqs. (5) and (7)
to eliminate $\mu_h$ and $\mu_Q$. 

So, finally, we have from Eqs. (5) - (8) all the chemical potentials in terms of
just one, $\mu_L$. We are now in a position to compute the ratio of $X$ to $B-L$ at
the end of era (b). Again assuming that the particles that carry $B$, $L$, and $X$ 
are light compared to $T_*$, one has

\begin{equation}
\begin{array}{ccl}
B & \propto & \frac{1}{3} \left( 18 \mu_Q - 9 \mu_{u^c} - 18 \mu_{d^c} + 9 \mu_{\overline{d^c}} \right) = 18 \mu_Q - 6 \mu_h = - 9 \mu_L, \\ & & \\
L & \propto & 12 \mu_L - 6 \mu_{\overline{L}} - 3 \mu_{\ell^c} =
21 \mu_L - 3 \mu_h = \frac{39}{2} \mu_L, \\ & & \\
X & \propto & 6 \mu_{\chi} - 6 \mu_{\chi^c} + (b(0)/f(0)) \mu_{\sigma} = 
12 \mu_{\chi} + 2 \mu_{\sigma} = - 9 \left( \frac{7p + 12}{p + 2} \right) \mu_L.
\end{array}
\end{equation} 

\noindent Consequently,

\begin{equation}
\frac{X}{B-L} = \frac{6}{19} \left( \frac{7p + 12}{p + 2} \right),
\end{equation}

\noindent which, by a very strange coincidence, is simply equal to 2 in the minimal case, where $p =1$. This is the ratio of $X$ to $B-L$ that 
exists also at the present era.

In order to obtain the present ratio of $X$ to $B$, which is our aim, we need
to consider what happened in era (c), when the present ratio of $B$ to $L$ was established. We assume that in era (c), where $T> T_W > M_W$, the electroweak symmetry is unbroken and therefore the chemical potential of the $W$ bosons vanishes and the chemical potentials are equal for all particles within any Standard Model multiplet.

In era (c), we no longer have to consider the quantum number $X$ or the chemical potentials of the $X$-bearing particles, as they do not affect the ratio of $B$ to $L$. The important chemical potentials are $\mu_Q$, $\mu_L$, and $\mu_h$. The 
chemical potentials of the other quarks and leptons are given in terms of these by the
relations in Eq. (4), which are still valid in era (c). Eq. (7), which gives the relation arising from the equilibrium of $SU(2)_L$ sphaleron processes, is also still valid.

The strategy is the same as the calculation done in era (b), but simpler. The first step is to use the condition that the universe has $Q=0$ to derive a formula for $\mu_h$ in terms of $\mu_Q$ and $\mu_L$. This relation is different from that for era (b), given in Eq. (5), because in era (b) the charge density included the contributions from the ``extra" quarks and leptons in $(5,1)$ and  $(\overline{5},2)$, namely the $\overline{\ell}$, $\overline{\nu}$, $\overline{d^c}$ and the half of the $\ell_{1,2}$, $\nu_{1,2}$ and $d^c_{1,2}$ with which they mate to obtain mass. Those fermions are light compared to $T$ in era (b); but in era (c) (or at least near the end of that era) we can neglect them because we assume that they are heavy compared to the electroweak scale and thus highly Boltzmann suppressed. 

Therefore the only particles that one must consider in computing the electric charge density are all the fermions of the Standard Model and the three electroweak Higgs doublets $h$, and $h'_{1,2}$. All these fermions may be treated as massless (since we are assuming that $SU(2)_L$ is unbroken in this era). However, the masses of $h$ and $h'_{1,2}$ must be taken into account.  
We therefore define the quantity
$c_h \equiv [b(m_h/T_W) + b(m_{h'_1}/T_W) + b(m_{h'_2}/T_W)]
/b(0)$.

Given all this, one has

\begin{equation}
\begin{array}{ccccl}
0 & = & Q & \propto  & 3 \mu_L (-1) + 3 \mu_{\ell^c} (+1)  \\ & & & & \\
& & & + & 9 \mu_Q (+ \frac{2}{3}) + 9 \mu_{u^c} (-\frac{2}{3}) 
+ 9 \mu_Q (- \frac{1}{3})
+ 9 \mu_{d^c} (+ \frac{1}{3}) \\ & & & & \\
& & & + & (b(0)/f(0)) c_h \mu_h (+1) \\ & & & & \\
& \Rightarrow & 0  &  = & - 6 \mu_L + 6 \mu_Q +(12 + 2 c_h) \mu_h
\\ & & & & \\
& \Rightarrow & 0  &  = & - 8 \mu_L  +(12 + 2 c_h) \mu_h
\;\;\; \Rightarrow  \;\; \mu_h  =  \frac{4}{6 + c_h} \mu_L,
\end{array}
\end{equation}

\noindent where we have used the $SU(2)_L$ sphaleron equilibrium condition $\mu_Q = - \frac{1}{3} \mu_L$, given in Eq. (7). Now that we have both $\mu_h$ and $\mu_Q$
in terms of $\mu_L$, we may compute the ratio of $B$ to $L$. Again, this gives a result different from Eq. (9), because of the different relation between temperature
and mass that holds in era (c). One obtains

\begin{equation}
\begin{array}{ccl}
B & \propto & \frac{1}{3} \left( 18 \mu_Q - 9  \mu_{u^c} - 9 \mu_{d^c}  \right) = 12 \mu_Q = - 4 \mu_L \\ & & \\
L & \propto & 6 \mu_L - 3 \mu_{\ell^c} =
9 \mu_L - 3 \mu_h = 9 \mu_L - 3 \frac{4}{6 + c_h} \mu_L = \frac{42 + 9 c_h}{6 + c_h} \mu_L, 
\end{array}
\end{equation} 

\noindent Therefore, when $T$ falls below $T_W$, the ratio $L/B$ is frozen at

\begin{equation} 
\frac{L}{B} = - \frac{3}{4} \left( \frac{14 + 3 c_h}{6 + c_h} \right) .
\end{equation}

\noindent Combining this with Eq. (10) gives

\begin{equation}
\frac{X}{B}  =  \frac{6}{19} \left( \frac{7p + 12}{p + 2} \right) \left(
\frac{66 + 13 c_h}{4(6 + c_h)} \right),
\end{equation}

\noindent which for the minimal case $p = 1$ reduces to

\begin{equation}
\frac{X}{B} = \frac{66 + 13c_h}{2(6 + c_h)}.
\end{equation} 

\noindent For the allowed range $0 < c_h < 3$ this varies between 5.5 and 5.833. If, as seems reasonable, one assumes that one linear combination of the three electroweak Higgs doublets (the ``Standard Model Higgs doublet") is much lighter than the others, one would expect $c_h \cong 1$, giving 
$X/B \cong 5.64$. A value of $X/B \approx 5.6$ implies that the dark matter particle has a mass close to 1 GeV. 

Besides the dark matter particle itself, there are five other flavors of $\chi(\chi^c)$ particles. These are, by definition, heavier than 
the dark matter particle and will all have decayed or annihilated by the time the temperature reaches $1$ GeV. It is important that the energy released in these decays and annihilations does not get trapped in the dark sector (i.e. the sector of $\chi$, $\chi^c$, $\sigma$, and $S$), as otherwise the thermal energy of the massless $S$ particles at the
time of primordial nucleosynthesis might cause the universe to expand too fast, leading to an excessive primordial Helium abundance. However, as long as $M_* <100$ TeV, the particles of the dark sector are kept in thermal contact with the Standard Model particles by $SU(2)_*$ gauge interactions and do not ``overheat".

In order for the dark matter to be almost purely asymmetric, there must be an efficient mechanism for dark matter particles and their antiparticles to annihilate into massless particles. This is why we introduced the massless $S$ fermion(s) and the scalar $\sigma$. 
Given the Yukawa coupling $y(\chi^c S) \sigma$, which was mentioned earlier, the exchange of a $\sigma$ allows the annihilation process $\chi^c + \overline{\chi^c} \longrightarrow S + \overline{S}$. In order to have the density of dark matter anti-particles very small compared to the density of dark matter particles, $m_{\sigma}/y$ must be less than about 10 GeV. Of course, this involves fine-tuning in the absence of supersymmetry or some other symmetry or mechanism that would make such a small scalar mass natural. 

In computing the ratios of $B$, $L$, and $X$ above, we made certain assumptions about the
$SU(2)_L$ and $SU(2)_*$ dynamics. In particular, we assumed that at the temperature when the
anomalous processes of one of these interactions become cosmologically negligible, the interaction in question may be treated as still unbroken. It is possible to make other assumptions \cite{HarveyTurner}. The result for the $B$, $L$, $X$ ratios would not greatly change. But to get an exact result one would need to understand the sphaleron dynamics and the details of the $SU(2)_L$ and $SU(2)_*$ phase transitions well.  

The whole scenario depends on there being a global charge $X$ that is conserved  
except for the $SU(2)_*$ anomaly (and possibly GUT-scale interactions). The question is why there should be such a global $U(1)_X$ and whether it is compatible with grand unification. The answer is that it can arise as an accidental symmetry of the low-energy theory. And, despite appearances, this can easily happen even in a grand unified version of this model. For example, consider an embedding of the model into $SU(5) \times SU(2)_*$. 
Suppose that all the Yukawa coupling allowed by $SU(5) \times SU(2)_*$ exist, 
except for $(\overline{5},2) (1,2) (5,1)_h$. (In other words, the following Yukawa couplings exist: 
$(10,1) (10,1) (5,1)_h$, $(10,1) (\overline{5}, 2) ( \overline{5}, 2)_h$,
$(5,1) (\overline{5}, 2) (1,2)_h$, and  $(1,2) (1,1) (1,2)_h$. This is easily ensured by a discrete global symmetry that commutes with $SU(5) \times SU(2)_*$. For instance, one can have a $Z_N$ symmetry under which $(10,1) \rightarrow \omega (10,1)$, $(5,1)_h \rightarrow \omega^{*2} (5,1)_h$, $(\overline{5},2)_h \rightarrow \omega^* (\overline{5}, 2)_h$, with all other multiplets transforming trivially. It is easy to show that with the coupling $(\overline{5},2) (1,2) (5,1)_h$ missing, the global $U(1)_X$ shown in Table I arises as an accidental symmetry of the low-energy theory. 

How would one observe dark matter in the laboratory? It would be very difficult to produce or detect it directly, since it interacts with the Standard Model particles only
by the $SU(2)_*$ gauge interactions, which are much more feeble than the weak interactions, because broken at a much higher scale. On the other hand, the ``extra" quarks and leptons that are in $(5,1)$ and $(\overline{5},2)$ could be directly produced in accelerators through their Standard Model interactions. These then could decay into a combination of Standard Model particles and the dark particles $\chi(\chi^c)$ by means of their $SU(2)_*$ gauge 
interactions. Each ``extra" fermion in $(\overline{5},2)$ is a partner in an $SU(2)_*$ doublet with a Standard model fermion, to which it can be converted by emitting an $SU(2)_*$ gauge boson. That boson, in turn, can decay into $\chi + \overline{\chi}$). The ``extra" fermions can also decay by ordinary charged weak interactions entirely into Standard Model particles. The point is that the ``extra" fermions mix slightly with the Standard Model fermions of the same color and charge. For example, 
the $\overline{d^c}$ mix with the left-handed $d$, $s$, and $b$ quarks with mixing angles that are of order $m_{d,s,b}/m_{\overline{d^c}}$, and similarly for the ``extra" leptons. 

The model described above is an illustration of a general idea that could be implemented in other ways. For example, one can construct an $SU(7)$ unified model that is in many ways quite similar to this. The fermions can be placed in three families, each consisting
of $21 + \overline{7} + \overline{7} 
+ \overline{7}$, which is the simplest way to incorporate a family in $SU(7)$. Under the $SU(5) \times SU(2)_*$ subgroup, each family decomposes into $(10,1) + (5,2) + (1,1) +
3 \times (\overline{5}, 1) + 3 \times (1,2)$. As in the model described earlier in this paper there would be $(1,2)$ fermions, which could be the dark matter, and ``extra fermions" in $5 + \overline{5}$ of the $SU(5)$. A difference with the model described earlier, which would be phenomenologically
significant, is that the extra fermions in the $SU(7)$ model would not be paired in $SU(2)_*$ doublets with ordinary Standard Model quarks and leptons. {\it Both} components of each $(5,2)$ get large mass with $(\overline{5},1)$ multiplets. Nevertheless, there would be
mixing between the ``extra fermions" and the Standard Model (SM) fermions. As in the model described earlier, those mixing angles would be of order the ratio of the masses of the SM fermions and extra fermions. The result would be that a heavier extra fermion would predominantly decay into a lighter extra fermion plus a dark matter pair, as its decays into a SM fermion plus dark matter pair would be suppressed by these small mixing angles. The {\it lightest} extra quark (or lepton) would have no choice, however, but to decay into SM quarks (or leptons). This would predominantly happen through the weak interactions, since, as in the model described earlier, the $\overline{d^c}$ would mix slightly with the left-handed $d$, $s$, and $b$, and similarly for the leptons. 

\section{Conclusions}

We have shown that it is possible to construct simple unified models in which the sphalerons of a new interaction convert asymmetries of $B$ and $L$ into a dark matter asymmetry. Since there are two kinds of sphaleron process involved, the equilibrium conditions allow one to compute the ratios of $B$, $L$, and $X$ (the dark matter number) independently of the nature of the primordial asymmetry, e.g. whether it was an asymmetry in $B$ or in $L$. Since one can compute the ratio of $X$ to $B$ in such models, one obtains a prediction for the mass of the dark matter particle. The dark matter particles in the scenario we describe does not have Standard Model gauge interactions and so would not be easily detectable in a direct way. However, such models generically give rise to extra vectorlike pairs of quarks and leptons that transform like $5 + \overline{5}$ of $SU(5)$. These could be directly produced, and decay into Standard Model fermions plus dark matter particle-antiparticle pairs. The phenomenology of such models remains to be explored. 

These models predict the number density of dark matter {\it a priori} but not their mass, leaving mass of the dark matter particle to be inferred from the measured dark matter density.
It would be interesting to see if a model could be constructed which predicts {\it a priori} 
both the number density and mass of the dark matter particles.

\section*{Acknowledgements} This work was
supported by U.S. DOE under contract DE-FG02-12ER41808.

\end{document}